\documentclass[twocolumn,preprintnumbers,superscriptaddress,amsmath,floatfix]{revtex4}
\usepackage{amssymb}  % \gtrsim, \geqslant, etc: see amsguide.ps
\usepackage{graphicx,subfigure}
\usepackage{exscale}
\usepackage{textcomp}
\usepackage{enumerate}
\usepackage{amsmath}
\usepackage{amssymb}
\usepackage{color,graphicx}
\usepackage[dvipsnames]{xcolor}
\usepackage{orcidlink}
\usepackage{color}

\usepackage[normalem]{ulem}  % \sout{old text} for strikeout

\newcommand{\chipt}{\chi_\mathrm{EFT}}
\newcommand{\nsat}{n_0}
\newcommand\si{\sigma}

\newcommand\om{\omega}
\newcommand{\be}{\begin{equation}}
\newcommand{\ee}{\end{equation}}
\newcommand{\bea}{\begin{eqnarray}}
\newcommand{\eea}{\end{eqnarray}}
\newcommand{\ba}[1]{\begin{array}{#1}}
\newcommand{\ea}{\end{array}}

\newcommand{\Msolar}{\ensuremath{{\rm M}_\odot}}

\newcommand{\sliver}{\kern 0.07em} % like italic correction \/

\newcommand{\MeV}{\text{MeV}}

\newenvironment{tightlist}[2]{ 
\begin{list}{#2}{
  \usecounter{#1}
  \setlength{\topsep}{0ex} % added to \parskip
  \setlength{\itemsep}{-\parsep} % added to \parsep
  \settowidth{\labelwidth}{#2} % width of item label
  \setlength{\labelsep}{0.2em}        % space between label and text
  \setlength{\leftmargin}{\labelwidth}% margin = width + separation
  \addtolength{\leftmargin}{\labelsep}
 }}{\end{list}}
\begin{document}

\title{Tabulated Equations of State From Models Informed by Chiral Effective Field Theory}

 \author{M.~G.~Alford \,\orcidlink{0000-0001-9675-7005}
}
 \email{alford@physics.wustl.edu}
 \affiliation{Department of Physics, Washington University in St.~Louis, St.~Louis, MO 63130, USA}
 
 \author{L.~Brodie\,\orcidlink{0000-0001-7708-2073
}}
 \email{b.liam@wustl.edu \text{(corresponding author)}}
 \affiliation{Department of Physics, Washington University in St.~Louis, St.~Louis, MO 63130, USA}
 
 \author{A.~Haber\,\orcidlink{0000-0002-5511-9565}}
 \email{ahaber@physics.wustl.edu}
 \affiliation{Department of Physics, Washington University in St.~Louis, St.~Louis, MO 63130, USA}

\author{I.~Tews\,\orcidlink{0000-0003-2656-6355
}}
\email{itews@lanl.gov}
\affiliation{Theoretical Division, Los Alamos National Laboratory, Los Alamos, NM 87545, USA}

\date{16 April 2023}

\preprint{LA-UR-23-23837}

\begin{abstract}
We construct four equation of state (EoS) tables, tabulated over a range of temperatures, densities, and charge fractions, relevant for neutron star applications such as simulations of neutron star mergers. The EoS are computed from a relativistic mean-field theory constrained by the pure neutron matter EoS from chiral effective field theory, inferred properties of isospin-symmetric nuclear matter, and astrophysical observations of neutron star structure. To model nuclear matter at low densities, we attach an EoS that models inhomogeneous nuclear matter at arbitrary temperatures and charge fractions. The four EoS tables we develop are available from the CompOSE EoS repository \href{https://compose.obspm.fr/eos/297}{compose.obspm.fr/eos/297} and \href{https://gitlab.com/ahaber/qmc-rmf-tables}{gitlab.com/ahaber/qmc-rmf-tables}.

\end{abstract}

\maketitle
\section{Introduction}
We developed a set of four relativistic mean-field theories (RMFTs) in Ref.~\cite{Alford:2022bpp}, which we call ``QMC-RMFs.'' These models are constrained by accurate quantum Monte Carlo (QMC)~\cite{Carlson:2014vla} calculations of neutron matter using chiral effective field theory ($\chipt$) interactions~\cite{Tews:2018kmu}, inferred properties of isospin-symmetric nuclear matter, and astrophysical observations of neutron star structure. The four QMC-RMFs from the $\chipt$ uncertainty band of Ref.~\cite{Tews:2018kmu} were chosen because they cover the full range of uncertainty in the binding energy around nuclear saturation density and the $2\sigma$ uncertainty in the mass of the heaviest known neutron star~\cite{Riley:2021pdl}. Other commonly used RMFTs, such as GM1~\cite{GM1}, NL3~\cite{Lalazissis:1996rd}, IU-FSU~\cite{Fattoyev:2010mx}, and SFHo~\cite{Steiner:2012rk}, are not informed by data for pure neutron matter. We expect our RMFTs will be of particular value to groups performing simulations of neutron star mergers or core-collapse supernovae. We also provide all necessary quantities to enable the study of transport properties in these systems. To enable these calculations, we tabulate all four models derived in Ref.~\cite{Alford:2022bpp} over a wide range of temperatures, densities, and charge fractions, such that they can be directly accessed by the simulation community. We provide detailed microscopic information in the equation of state (EoS) tables that allow the user to extract nucleon dispersion relations and effective nucleon masses. 

This paper builds on the work in Ref.~\cite{Alford:2022bpp} by extending our investigation to the thermodynamic environments relevant to neutron star mergers in the following ways:
\newcounter{counta}
\begin{tightlist}{counta}{$\bullet$}
\item Since the proton to total baryon fraction (charge fraction) in a neutron-star merger is typically less than 25 percent, an accurate description of neutron-rich nuclear matter is essential for neutron star (merger) physics. We achieve this by initially fitting our models to $\chipt$ calculations of neutron matter \cite{Tews:2018kmu} at $T=0$ MeV in Ref.~\cite{Alford:2022bpp}. In this work, we show consistency with finite temperature $\chipt$ calculations at $T=20$ MeV \cite{Keller:2020qhx}. \item We describe in detail a thermodynamically consistent procedure to attach a low-density EoS that models inhomogeneous nuclear matter in and out of chemical equilibrium. \item We show the significance of antibaryons on the EoS in the temperature range potentially relevant for neutron star mergers. 
  \end{tightlist}  

In a neutron star merger, matter is compressed and rarefied across a wide range of densities. Our QMC-RMFs describe the behavior of homogeneous nuclear matter, which is a valid description of nuclear matter down to densities of about $0.5\nsat$, where $\nsat=0.16\,\text{fm}^{-3}$ is nuclear saturation density. 
To accurately describe matter, at finite temperatures and not necessarily in chemical equilibrium, at densities much less than $0.5\nsat$, we use the Hempel and Schaffner-Bielich statistical model of nuclear matter \cite{Hempel:2009mc} applied to the IU-FSU (IUF for short) RMFT \cite{Fattoyev:2010mx}, called HS(IUF). 
We showed, in Ref.~\cite{Alford:2022bpp}, that the IUF RMFT model of homogeneous nuclear matter violates $\chipt$ constraints on the binding energy per nucleon in pure neutron matter at zero temperature. 
The EoS tables presented here combine the best of both approaches by attaching HS(IUF) to a QMC-RMF through a phase transition, explained in Sec.~\ref{sec: attach_phases}. HS(IUF) accurately models inhomogeneous nuclear matter and our QMC-RMFs accurately model homogeneous nuclear matter. 
 Whenever the proton chemical potential is small compared to the temperature, anti-nucleons give a non-negligible contribution to the EoS, but are often neglected. 
Consequently, we include interacting anti-nucleons and anti-protons in our calculations.

We host our EoS tables on the CompOSE website \href{https://compose.obspm.fr/eos/297}{compose.obspm.fr/eos/297} \cite{Typel:2013rza,qmc_rmf1,qmc_rmf2,qmc_rmf3,qmc_rmf4} and as ``hdf5" tables on GitLab \href{https://gitlab.com/ahaber}{gitlab.com/ahaber}. 
CompOSE, short for CompStar Online Supernovae Equations of State \cite{Typel:2013rza}, is an online repository and a compiler that provides thermodynamic and microphysical properties of different EoSs, allows for interpolation between data points, and can calculate additional thermodynamic quantities beyond those provided in the uploaded EoS files.

The rest of this paper is organized as follows. 
In Sec.~\ref{sec:models}, we discuss how different thermodynamic properties can be derived from an RMFT, which we use to model homogeneous nuclear matter. We also discuss the EoS we chose to model the low-density behavior of nuclear matter. In Sec.~\ref{sec: attach_phases}, we provide the steps that were used to attach the low-density and high-density phases of nuclear matter. In Sec.~\ref{sec:results}, we discuss the format of the EoS tables and what quantities are provided. We also compare some properties of our EoSs to observational and experimental constraints. In all our calculations, we use natural units, \mbox{$\hbar=c=k_B=1$}.

\section{Microscopic Models}
\label{sec:models}
\subsection{Modeling the High-Density Phase}

\begin{figure}
    \centering{
    \includegraphics[height=9cm,width=1\columnwidth]{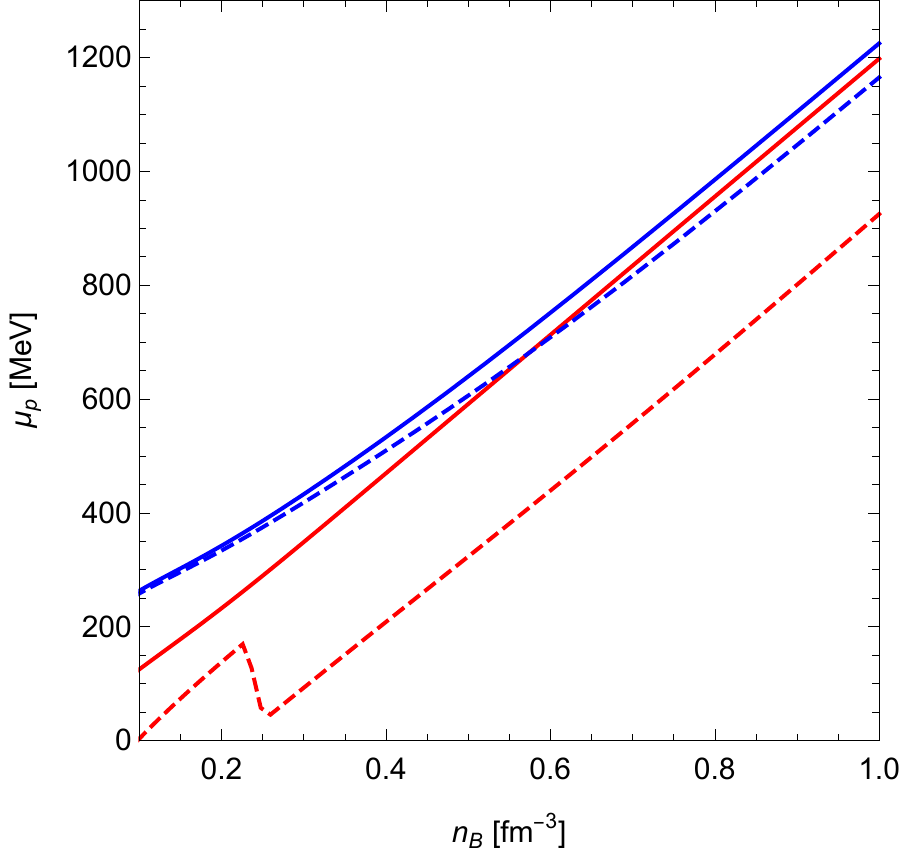}
    }
    \caption{Proton chemical potential $\mu_p$ as a function of baryon density at $T=100$ MeV (blue lines) and $T=150$ MeV (red lines) with (solid lines) and without (dashed lines) anti-nucleons for QMC-RMF3 at $y_q=0.01$. Note that
    the dashed red line shows an apparent phase transition at high densities (visible as a kink) resulting from the incorrect comparison 
    between the QMC-RMF phase without anti-nucleons and the HS(IUF) model which includes anti-nucleons.}
    \label{fig:anti}
\end{figure}

To model homogeneous nuclear matter over a wide range of temperatures and charge fractions ($0\leq T \leq 150$ MeV and $0.01 \leq y_q \leq 0.6$), we use an RMFT. 
An RMFT models the behavior of neutrons and protons interacting via a mesonic mean-field background; we include $\sigma$, $\omega$, and $\rho$ mesons. 
We also include neutrons, protons, electrons, positrons, and photons -- the same particles as in Ref.~\cite{Alford:2022bpp} -- with the addition of anti-neutrons and anti-protons which become significant at high $T$ and low $y_q$, because the particle chemical potentials are small. The addition of anti-particles, while having little influence on the pressure and the energy density of the system, can change the chemical potentials significantly. This change in the chemical potentials will influence transport properties and weak-interaction processes like modified and direct Urca, whose rates were studied in Refs.~\cite{Alford:2021ogv,Most:2022yhe,Alford:2022ufz}. We compare the proton chemical potential with anti-nucleons (solid lines) and neglecting anti-nucleons (dashed lines) in Fig.~\ref{fig:anti} at two different temperatures as a function of density for QMC-RMF3 at $y_q=0.01$.
Neglecting anti-nucleons leads to a very late phase transition from the low-density to the high-density phase (visible as a kink in the red, dashed line) because HS(IUF) does include anti-nucleons. 

Muons could be added to the theory as another non-interacting lepton field, but they play a
sub-leading role compared to the electrons in establishing electrical neutrality of bulk nuclear matter and affect the total pressure at the one percent level. The proton fraction can change from roughly 10$\%$ up to 15$\%$
compared to the case without muons however, the smaller Fermi
momentum of the electrons counterbalances this effect.
The resulting effect on the direct Urca threshold density
is therefore only at the few percent level. We find that none of our models possess a direct Urca threshold at the relevant densities for neutron star physics, independent of the inclusion of muons.

The coupling constants of the four QMC-RMFs developed in Ref.~\cite{Alford:2022bpp}, and selected properties are displayed in Table~\ref{tab:couplings}.

\begin{table*}
\begin{tabular}{lccccccccccc}
\hline
 & $g_\si$ & $g_\om$ & $g_\rho$ & $b$ & $c$ & $b_1$ & $B$\\[-0.3ex]
     &     &     &     &     &     &   & [MeV$^4$]\\ 
\hline
QMC-RMF1 & 7.54 & 8.43 & 10.88 & 0.0073 & 0.0035 & 7.89 & -577307\\
QMC-RMF2 & 7.82 & 8.99 & 11.24 & 0.0063 & -0.0009 & 8.02 & -460210\\
QMC-RMF3 & 8.32 & 9.76 & 11.02 & 0.0063 & -0.006 & 5.87 & -723934\\
QMC-RMF4 & 8.21 & 9.94 & 12.18 & 0.0041 & -0.0021 & 10.43 &  -150000\\
\hline
\end{tabular}
\caption{Parameters describing the QMC-RMFs used for the high-density phase of nuclear matter from Ref.~\cite{Alford:2022bpp}. All QMC-RMFs use the same mass for the proton and neutron, $m_N=939\,\MeV$; the electron mass is $m_e=0.511\,\MeV$; the meson masses are $m_{\sigma} = 491.5\,\MeV$, $m_{\omega} = 782.5\,\MeV$, and $m_{\rho} = 763.0\,\MeV$.
}
\label{tab:couplings}
\end{table*}

In the remainder of this section, we describe the calculation of the pressure, from which all thermodynamic quantities can be derived. The pressure is a sum of partial pressures from nucleons, mesons, leptons, photons, and a pressure offset $B$ that allows us to have some control over where the phase transition to an inhomogeneous phase of matter occurs,
\begin{equation}
\label{eq:total_pressure}
    P=P_{N} + P_{M} + P_{l} + P_{\gamma} + B \ .
\end{equation}
The nucleonic partial pressure is found by neglecting the quantum fluctuations of the meson fields (the mean-field approximation) and by assuming the mean fields are spatially and temporally uniform. One can then calculate the one-loop contribution from the nucleon fields (see Sec.~3.1 of  Ref.~\cite{Schmitt:2010pn}). The resulting partial pressure is that of a free-fermion gas with a meson-modified dispersion relation $E_i$ 
\begin{equation}
     P_N=2T\sum_{i=n,p,\bar{n},\bar{p}}\int\frac{d^3k}{(2\pi)^3}\ln\left[1+e^{-\left( E_i-\mu_i\right)/T}\right] \, .
 \end{equation}
The nucleon partial pressure is a sum of neutron, proton, anti-neutron, and anti-proton contributions, where $T$ is the temperature and $k$ is the momentum integrated over. The neutron and proton chemical potentials are $\mu_n$ and $\mu_p$. The anti-neutron and anti-proton chemical potentials are $\mu_{\bar{n}}$ and $\mu_{\bar{p}}$, where $\mu_{\bar{n}}=-\mu_n$ and $\mu_{\bar{p}}=-\mu_p$. $E_i$ is the nucleon dispersion relation
\begin{equation}
\label{eq:nuc_dispersion_relation}
    E_i=\sqrt{k^2+M_*^2} - U_i \, ,
\end{equation}
where 
\begin{equation}
\label{eq:vector_offset}
    U_i = -g_{\omega}\langle\omega_0\rangle-g_{\rho}I_{3i}\langle\rho_{03}\rangle \, .
\end{equation}
The energy shift in the dispersion relation is due to the interaction of nucleons with vector mesons, where $g_{\omega}$ and $g_{\rho}$ are the $\omega$ and $\rho$ nucleon-meson Yukawa couplings. 
The angled brackets around the meson fields indicate that the mean-field approximation has been used; see Sec.~IIB of Ref.~\cite{Alford:2022bpp} for details about why only certain components of the meson fields remain in the mean-field approximation. 
$I_{3i}$ is the third component of the isovector of isospin generators, where $I_{3n}=-1/2$, $I_{3p}=+1/2$, $I_{3\bar{n}}=+1/2$, and $I_{3\bar{p}}=-1/2$. The relativistic Dirac effective nucleon mass $M_*$ is 
\begin{equation}
\label{eq:dirac_eff_mass}
    M_* = m_N - g_{\sigma}\langle\sigma\rangle\, ,
\end{equation}
where $g_{\sigma}$ is the nucleon--$\sigma$-meson coupling constant. 
The bare nucleon mass $m_N=939\,\MeV$, which we use for both neutrons and protons, is modified by the $\sigma$ field to capture the effects of interactions in the nuclear medium. 
Note that Eq.~\ref{eq:dirac_eff_mass} is not equivalent to the Landau effective mass
\begin{equation}
\label{eq:landau_eff_mass}
    m^L_{i}=\sqrt{k_{Fi}^2+M_*^2} \, ,
\end{equation}
which yields different values for neutrons and protons even with the same bare masses.
Here, we extend the definition of Fermi momentum to finite temperatures via
\begin{equation}
    k_{Fi}=\left(3\pi^2 n_{i}\right)^{1/3} \,
\end{equation}
with $n_i$ being the net (particle minus anti-particle) number density for each particle species.
In the mean-field approximation, the mesonic partial pressure is just the mesonic portion of the Lagrangian in the mean-field approximation,
\begin{align}
    P_M =& 
    -\frac{1}{2}m^2_{\sigma}\langle\sigma\rangle^2
    -\frac{bM}{3}(g_{\sigma}\langle\sigma\rangle)^3-\frac{c}{4}(g_{\sigma}\langle\sigma\rangle)^4  \\
    & +\frac{1}{2}m^2_{\omega}\langle\omega_0\rangle^{2}  +\frac{1}{2}m^2_{\rho}\langle\rho_{03}\rangle^2+b_{1}(g_{\rho}\langle\omega_0\rangle\langle\rho_{03}\rangle)^2\,,\nonumber
\label{eq:mesons}
\end{align}
where the meson masses are $m_{\sigma} = 491.5\,\MeV$, \mbox{$m_{\omega} = 782.5\,\MeV$}, and $m_{\rho} = 763.0\,\MeV$.
The meson self-interaction strength parameters $b$, $M$, $c$, $b_1$ are fit (along with $g_{\sigma}$, $g_{\omega}$, and $g_{\rho}$) following the procedure described in Ref.~\cite{Alford:2022bpp} to reproduce nuclear saturation properties, maintain consistency with $\chipt$ calculations of neutron matter, and meet astrophysical constraints (see Table~\ref{tab:couplings} for values of these parameters). 

The lepton partial pressure is a sum of electron and positron contributions,
 \begin{equation}
     P_e= 2T\sum_{j=\pm1}\int\frac{d^3k}{(2\pi)^3}\ln\left[1+e^{-\left( \sqrt{k^2+m_e^2}-j\mu_e\right)/T}\right] \, ,
 \end{equation}
where $m_e = 0.511\,\MeV$, $j=1$ represents an electron, and $j=-1$ a positron. The free photon partial pressure is 
\begin{equation}
    P_{\gamma} = \frac{\pi^2}{45}T^4\, .
\end{equation}

The pressure offset $B$ is fixed for each QMC-RMF following the procedure described in Sec.~\ref{sec: attach_phases}.

Numerical values of the total pressure $P$ are calculated in the following self-consistent way. For a given set of coupling constants at given temperature and chemical potentials (one can alternatively fix temperatures, number densities, and charge fractions), the mesonic mean fields are found by simultaneously solving the equations of motion
\begin{align}
\label{eq:EoM_pressure_extremum}
\frac{\partial P}{\partial\langle\sigma\rangle}=0 \, , \qquad \frac{\partial P}{\partial\langle\omega_0\rangle}=0 \, , \qquad \frac{\partial P}{\partial\langle\rho_{03}\rangle}=0 \,, 
\end{align}
along with the condition that the number density of protons and electrons are equal (charge neutrality)
\begin{equation}
\label{eq:EoM_charge_neutrality}
    n_p - n_e =0 \ ,
\end{equation}
and that the total number density of baryons is equal to the sum of neutron and proton number densities
\begin{equation}
\label{eq:EoM_baryon_number}
    n_B - n_n - n_p =0 \ .
\end{equation}
Note that these are net number densities, i.e., the difference between particle and anti-particle number densities.

The entropy density can be found by taking a derivative of the total pressure at a fixed value of the chemical potentials with respect to the temperature 
\begin{equation}
    s = \frac{\partial P}{\partial T}\bigg\vert_{\mu_i}.
\end{equation}
In the mean-field approximation, the mesonic pressure has no explicit temperature dependence; therefore, the mesons do not contribute to the entropy density. The photonic contribution is given by
\begin{equation}
    s_\gamma=\frac{4\pi^2}{45}T^3 \, .
\end{equation}
To calculate the energy density $\epsilon$, one needs to either use the thermodynamic relation
\begin{equation}
\label{eq:thermo_relation}
    \epsilon = -P + Ts + \sum_i \mu_i n_i
\end{equation}
or calculate it directly. To calculate the energy density directly it is useful to break it into the following sum
\begin{equation}
\label{eq:energy_density}
    \epsilon = \epsilon_N + \epsilon_M + \epsilon_l + \epsilon_{\gamma} - B\, .
\end{equation}
The nucleonic and lepton contributions are the standard sums over Fermi-Dirac functions and particle dispersion relations
\begin{equation}\label{eq:nuc_e}
     \epsilon_N = \sum_{i=n,p,\bar{n},\bar{p}}\int\frac{d^3k}{(2\pi)^3}E^*_i f_{Ni} \, ,
 \end{equation}
where 
\begin{equation}\label{eq:eeff}
    E^*_i = \sqrt{k^2+M_*^2} \, ,
\end{equation}
and the Fermi-Dirac function $f_{Ni}$ for nucleons is 
\begin{equation}
    f_{Ni} = \frac{1}{1+e^{(E_i-\mu_i)/T}}\, .
\end{equation}
Note that Eq.~(\ref{eq:nuc_e}) does not depend on the energy shift $U$; for a derivation see Ref.~\cite{Glendenning1996}.
The lepton energy density is a sum of electron and positron contributions
\begin{equation}
     \epsilon_l = \sum_{j=\pm 1}\int\frac{d^3k}{(2\pi)^3}\sqrt{k^2+m_e^2} f_{lj} \, ,
 \end{equation}
where
\begin{equation}
    f_{lj} = \frac{1}{1+e^{(\sqrt{k^2+m_e^2}-j\mu_j)/T}}\, .
\end{equation}
The photon contribution is 
\begin{equation}
    \epsilon_{\gamma} = \frac{\pi^2}{15} T^4\, .
\end{equation}
The meson contribution can be computed from the $T_{00}$ component of the stress-energy tensor $T^{\mu \nu}$ (see page 34 of Ref.~\cite{Schmitt:2014eka} for an example), where
\begin{equation}
    T^{\mu \nu} = 2\frac{\delta \mathcal{L}}{\delta g_{\mu \nu}}-g^{\mu \nu}\mathcal{L}\, .
\end{equation}
Using the Lagrangian from Ref.~\cite{Alford:2022bpp}, the energy density for the mesons is the following 
\begin{align}
\epsilon_M &=  \frac{1}{2}m_{\sigma}^2\langle\sigma\rangle^2 + \frac{1}{3} b m_N (g_{\sigma}\langle\sigma\rangle)^3 + \frac{c}{4}(g_{\sigma}\langle\sigma\rangle)^4
\\
&+ \frac{1}{2}m_{\omega}^2\langle\omega_0\rangle^2 + \frac{1}{2} m_{\rho}^2\langle\rho_{03}\rangle^2 + 3 b_1 (g_{\rho}\langle\omega_0\rangle\langle\rho_{03}\rangle)^2\, .\nonumber
\end{align}

\subsection{Modeling the Low-Density Phase}
Our QMC-RMFs assume that nuclear matter is homogeneous, which is a valid assumption down to about $0.5\nsat$. 
At low densities, nucleons begin to cluster together into nuclei so we need a description of inhomogeneous nuclear matter at the physically relevant range of temperatures and charge fractions. 
We use the HS(IUF) EoS \cite{Hempel:2009mc} to model this low-density physics.
HS(IUF) is the Hempel and Schaffner-Bielich nuclear statistical equilibrium model applied to the IUF RMFT. The HS(IUF) EoS is able to model finite nuclei and the density-driven phase transition to homogeneous nuclear matter via excluded volume effects. However IUF is inconsistent with $\chipt$ constraints on the binding energy per nucleon for pure neutron matter above $\approx \nsat$ \cite{Alford:2022bpp}, which is why we impose a phase transition to our QMC-RMFs at subsaturation densities when $y_q$ is close to zero.

\section{Attaching the low-density and high-density phase}
\label{sec: attach_phases}
We join the HS(IUF) phase of inhomogeneous nuclear matter to the QMC-RMF phase of homogeneous nuclear matter at a thermodynamically consistent phase transition.
The process is the following:

\begin{itemize}
    \item Since the HS(IUF) EoS is tabulated with ($T$, $n_B$, $y_q$) as independent variables, we solve the mean-field equations (Eqs.~\ref{eq:EoM_pressure_extremum}, \ref{eq:EoM_charge_neutrality}, \ref{eq:EoM_baryon_number}) for each of our four QMC-RMFs at each ($T$, $n_B$, $y_q$) point on the HS(IUF) EoS grid in the range: $T\in [0.1,158.48]$ MeV, $n_B\in [10^{-12},1.58 \text{ fm}^{-3}]$, $y_q\in [0.01,0.6]$. For exact values of the gridpoints, see the files \emph{eos.t}, \emph{eos.nb}, and \emph{eos.yq} in Refs.~\cite{qmc_rmf1,qmc_rmf2,qmc_rmf3,qmc_rmf4}. 
    \item The location of the phase transition depends on $B$, a pressure offset in the QMC-RMF pressure (Eq.~\ref{eq:total_pressure}). For each QMC-RMF, we fix $B$ by requiring that at low temperatures ($T\approx 0.1\,\MeV$) the phase transition from the HS(IUF) inhomogeneous phase to the QMC-RMF homogeneous phase occurs at densities below $\approx 0.5 n_0$ for low charge fractions ($y_q\lesssim 0.1$). This choice is made because the QMC-RMFs are constrained by $\chipt$ between $0.5\nsat$ and $2\nsat$ at $y_q = 0$ (pure neutron matter) and zero temperature. 
   
    \item For each $T$ and $y_q$ slice through the three-dimensional ($T,n_B,y_q$) EoS grid, we compare the pressure in each phase at fixed chemical potential; the phase with the greater pressure is the thermodynamically favored one. The only chemical potential that is equal between two phases of charge-neutral nuclear matter at fixed temperature and charge fraction is (see case Ic in Table II in Ref.~\cite{Hempel:2009vp} for more details) 
    \begin{equation}
    \label{eq:mu_z}
        \mu_z = (1-y_q) \mu_n + y_q (\mu_p + \mu_e)\ ,
    \end{equation}
    which can be thought of as the chemical potential for a ``$z$'' particle (containing $y_q$ of a proton, $y_q$ of an electron, and $1-y_q$ of a neutron) that diffuses between the two phases of matter to maintain fixed $y_q$, $T$, and charge neutrality on both sides of the phase boundary. To compare pressures, we compute $P(\mu_z)$ in the QMC-RMF homogeneous phase at fixed $T$ and $y_q$. Next, we find the transition point $\mu_z^\text{trans}$ where the QMC-RMF pressure is equal to the HS(IUF) pressure. We then attach the two phases by using the tabulated HS(IUF) EoS from the CompOSE database \cite{compose:iuf} up to the transition and the QMC-RMF EoS above it. 
    \item At the phase transition chemical potential $\mu_z^{\text{trans}}$, the slope of the pressure changes; therefore, $n_B$ jumps discontinuously, since \mbox{$\partial P/\partial \mu_z |_{T,y_q} = n_z = n_B$}, where $n_z=n_B$ by the definition of $\mu_z$. In the density gap between the two phases, there exists a mixture of each phase, where the properties of each phase remain the same, only the volume fraction changes. To construct a mixed phase, we perform a weighted volume average of the thermodynamic properties between the two pure phases as a function of $n_B$ at fixed $T$ and $y_q$. 
\end{itemize}

\section{The Resulting Equations of State}
\label{sec:results}

\subsection{Format and Content}
We provide an EoS table for each of our four QMC-RMFs. 
Each EoS table contains $T,n_B,y_q$ as independent variables. 
We provide thermodynamic properties of the EoS, particle composition, microphysical properties, and stellar predictions of the EoS. 
We host our EoS tables on CompOSE \href{https://compose.obspm.fr/eos/297}{compose.obspm.fr/eos/297} \cite{qmc_rmf1,qmc_rmf2,qmc_rmf3,qmc_rmf4} and GitLab \href{https://gitlab.com/ahaber/qmc-rmf-tables}{gitlab.com/ahaber/qmc-rmf-tables}. 
For details about how to extract these quantities from the CompOSE tables, see the manual \cite{CompOSECoreTeam:2022ddl} or the quick guide for providers and users \cite{compose_quick_guide}.
We provide the following thermodynamic properties for all four QMC-RMF tables: pressure, entropy density, all particle chemical potentials, free energy density, energy density, and $\mu_z$ (Eq.~\ref{eq:mu_z}).
The pressure and energy density for the QMC-RMF phase is computed using Eq.~\ref{eq:total_pressure} and Eq.~\ref{eq:energy_density}; see Sec.~IID in Ref.~\cite{Hempel:2009mc} for details about how the pressure and energy density were calculated in the HS(IUF) EoS. The entropy density is related to other thermodynamic quantities by Eq.~\ref{eq:thermo_relation}. $\mu_z$ is defined in Eq.~\ref{eq:mu_z}.
The free energy density is defined as
    \begin{equation}
        f = P - Ts\, .
    \end{equation} 

Additionally, we provide all models in cold $\beta$-equilibrium, where $\mu_n = \mu_p + \mu_e$, as a function of temperature and density \cite{compose:QMC-RMF3cold,Alford:2022bpp}. Note that we specify \emph{cold} $\beta$-equilibrium because at temperatures $T\gtrsim 1\,\MeV$ finite-temperature corrections to this expression become non-negligible \cite{Alford:2018lhf,Alford:2021ogv}.

The particle composition, microscopic quantities, and stellar predictions of the EoS are provided whenever they are available. For the particle composition, we list the following information: a phase index where ``phase 1'' is the HS(IUF) phase, the EoS we use to model low-density nuclear matter, ``phase 2'' is a volume-averaged mixture of phase 1 and phase 3, and ``phase 3'' is a QMC-RMF; the ratio of the number of neutrons, protons, electrons, and number of $^2$H, $^3$H, $^3$He, $^4$He nuclei to the number of baryons; the average mass, average charge number, and total charge fraction for a specified group of nuclei. 
We also list the Dirac effective nucleon mass (Eq.~\ref{eq:dirac_eff_mass}), the energy shift $U_i$ in the nucleon dispersion relation (Eq.~\ref{eq:nuc_dispersion_relation}) due to the interaction of nucleons with vector mesons (Eq.~\ref{eq:vector_offset}), and the Landau effective nucleon mass (Eq.~\ref{eq:landau_eff_mass}). The Landau effective mass is not available in phase 1, the HS(IUF) phase. The nucleon dispersion relation energy shift $U_i$ is only available in phase 1 when $n_B>\nsat$ and always available in phase 3. 
The predicted masses and radii of neutron stars in $\beta$-equilibrium at zero temperature are obtained by solving the Tolman-Oppenheimer-Volkoff (TOV) equations.

\subsection{Properties of the Equations of State}
The four EoSs we have tabulated are consistent with current astrophysical observations and experimental constraints. They are consistent with $\chipt$, nuclear saturation properties, and observations of neutron star structure.

$\chipt$ is a low-energy effective theory of quantum chromodynamics.
Its coupling parameters are fitted to nucleon-nucleon scattering data and binding energies of atomic nuclei~\cite{Epelbaum:2008ga,Machleidt:2011zz}. 
Using QMC methods, it has been used to calculate the binding energy per nucleon of pure neutron matter ($y_q=0$) from about $0.5\nsat$ to $2\nsat$ at zero temperature~\cite{Tews:2018kmu}, which provides constraints for our models. 
In our EoS tables, when $y_q$ is close to zero, the QMC-RMF phase is favored over the HS(IUF) phase, meaning the pressure is higher at the same $\mu_z$. The QMC-RMFs were developed by satisfying the binding energy constraint from $\chipt$ (see Fig.~2 in Ref.~\cite{Alford:2022bpp}) so this constraint is met in our EoS tables. Note that the choice of the pressure offset $B$ only changes the definition of the vacuum pressure and does not affect a physical observable such as the binding energy. There is no effect on the binding energy because the binding energy is the difference between the energy density of the system and the rest energy density of the system without particle interactions
\begin{equation}
\label{eq:binding_energy}
    \mathcal{E} = \epsilon - (m_N n_B - B)\, ,
\end{equation}
where the $B$ term in Eq.~\ref{eq:binding_energy} cancels the $-B$ in Eq.~\ref{eq:energy_density}. 
\begin{figure}
    \centering{
    \includegraphics[height=9cm,width=1\columnwidth]{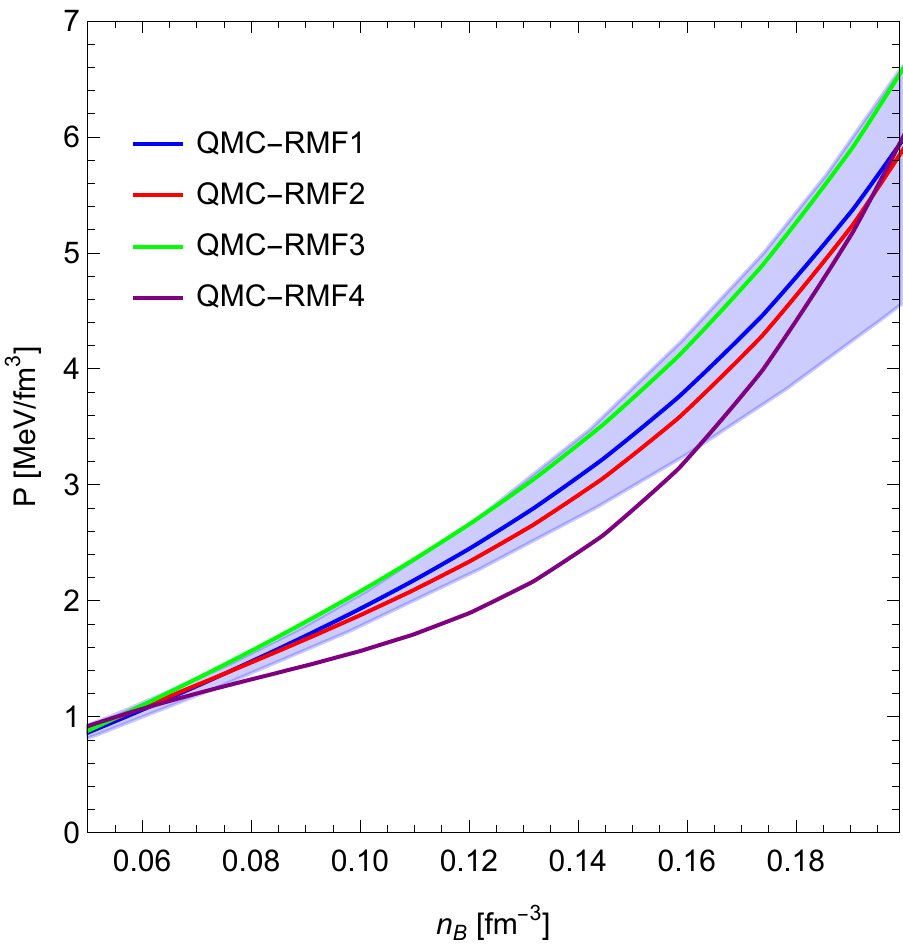}
    }
    \caption{Pressure as a function of baryon number density in neutron matter at $T=20$ MeV. We compare our QMC-RMFs (solid lines), which were calibrated to $T=0$ properties of neutron matter using the $\chipt$ calculation of Ref.~\cite{Tews:2018kmu}, to the finite temperature $\chipt$ calculation of Ref.~\cite{Keller:2020qhx} (blue shading). All QMC-RMFs are consistent with Ref.~\cite{Tews:2018kmu}, and QMC-RMF1,2,3 are consistent with the finite temperature calculation shown here.
    }
    \label{fig:20mev_chieft_comparison}
\end{figure}
To date, there are no $\chipt$ calculations of the properties of neutron matter at finite temperature using QMC methods. Using many-body perturbation theory, however, the properties of neutron matter have been calculated at finite temperatures. We compare our QMC-RMFs to the $\chipt$ calculation of Ref.~\cite{Keller:2020qhx} at $T=20$ MeV in Fig.~\ref{fig:20mev_chieft_comparison}, which uses a less conservative estimate for the systematic errors than Ref.~\cite{Tews:2018kmu}. We find that QMC-RMF1,2,3 are consistent with the finite temperature calculation but not QMC-RMF4.

The nuclear saturation properties of the EoSs, given in terms  of the parameters  $B_{\text{sat}}$, $\kappa$, $J$, and $L$ are found by expanding the binding energy per nucleon both around $\nsat$ and $y_q=0.5$ at $T=0$,
\begin{equation}
    \mathcal{E}(n_B,\alpha)=(B_\text{sat}+\frac{\kappa}{2!}\delta^2+\cdots)+\alpha^{2}(J+L\delta+\cdots)+\cdots,
    \label{eq:snm_power_series}
\end{equation}
where $\delta\equiv(n_{B}-\nsat)/(3\nsat)$ and $\alpha\equiv(n_n-n_p)/(n_n+n_p)=1-2 y_q$; $\alpha$ represents the asymmetry of nuclear matter, where $n_n$ and $n_p$ are the neutron and proton number densities, respectively.
In isospin-symmetric nuclear matter, $\alpha=0$. $B_\text{sat}$ is the binding energy at saturation density, $\kappa$ is the incompressibility of nuclear matter, $J$ is the  symmetry energy at $\nsat$ in the quadratic approximation,
\begin{equation}
    J = \frac{1}{2}\frac{\partial^2 \mathcal{E}(n_B,\alpha)}{\partial \alpha^2}\bigg|_{\nsat,\alpha=0}\, ,
\end{equation}
and $L$ is the slope of the symmetry energy at $\nsat$
\begin{equation}
    L = 3n_B \frac{\partial}{\partial n_B}\bigg(\frac{1}{2}\frac{\partial^2 \mathcal{E}(n_B,\alpha)}{\partial \alpha^2}\bigg)\bigg|_{\nsat,\alpha=0}\, .
\end{equation}
Around $\nsat$ and $y_q=0.5$ at $T=0$, the HS(IUF) phase is favored over the QMC-RMF phase. Both phases produce similar properties of nuclear matter around saturation density (see Table~I and II in Ref.~\cite{Alford:2022bpp}). The coupling constants of the IUF RMFT were found by fitting to properties of finite nuclei, and the QMC-RMF couplings were found by fitting to properties of pure neutron matter and bulk properties of isospin-symmetric nuclear matter. Therefore, each phase is favored in our tables in the parameter range where each phase is expected to be reliable. The properties of nuclear matter at saturation density from each of our EoS tables are the same: $\nsat=0.1546$ fm$^{-3}$, $B_{\text{sat}}=-16.39\,\MeV$, $\kappa=231.3\,\MeV$, $J=31.29\,\MeV$, and $L=47.20\,\MeV$. These values are consistent with the white intersection region in Ref.~\cite{Drischler:2020hwi} and with Refs.~\cite{Shlomo:2006incompressibility,Horowitz:2020evx}. There is a strong correlation between $L$ and the neutron skin for RMFTs that have the same interaction terms as our QMC-RMFs \cite{Reed:2023cap}. Since our QMC-RMFs are consistent with inferred constraints on $L$, it is safe to assume our models predict reasonable values for the neutron skin. Another value of interest at saturation density is the Dirac effective nucleon mass (Eq.~\ref{eq:dirac_eff_mass}), which we report in Table~\ref{tab:properties}. This effective mass is calculated in $\beta$-equilibrium at zero temperature and is consistent with values found by other microscopic calculations \cite{Bodmer:1989hdx,Margueron:2017eqc,Li:2018lpy,Chen:2014mza}.

\begin{figure}
    \centering{
    \includegraphics[height=9cm,width=1\columnwidth]{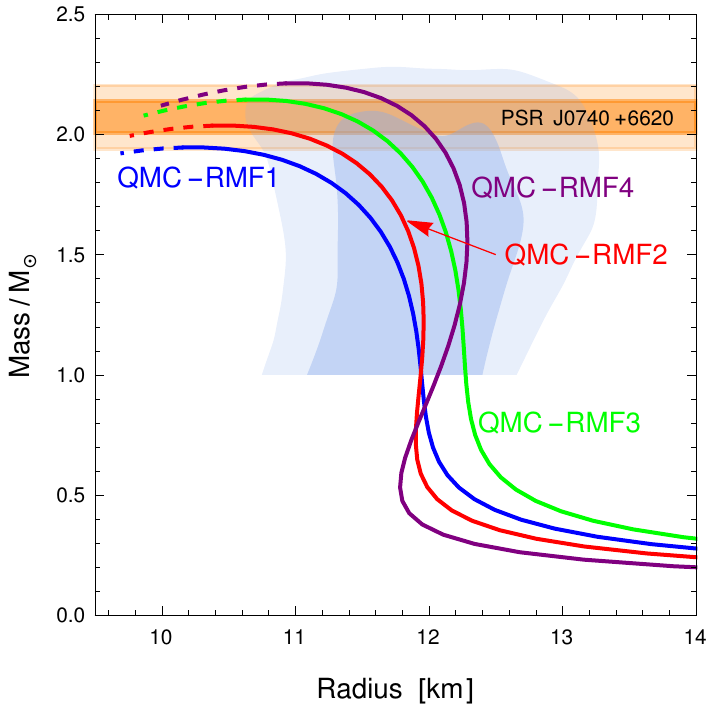}
    }
    \caption{Mass-radius curves of the four EoSs we developed. These four mass-radius curves span the sample acquired by constraining RMFTs with the method described in Ref.~\cite{Alford:2022bpp}. The HS(IUF) $\beta$-equilibrated inhomogeneous phase is present at baryon densities below $\approx$ $0.3\nsat$ (see Table~\ref{tab:properties}). The orange-shaded bars show the $68\,\%$ credibility (dark shading) and $95\,\%$ credibility (light shading) mass measurement of pulsar J0740+6620 from Ref.~\cite{Riley:2021pdl}. 
    The blue-shaded area shows the 68\,\% credibility (dark shading) and 95\,\% credibility (light shading) multi-messenger constraints from Ref.~\cite{Pang:2021jta}.
    }
    \label{fig:mr_curve}
\end{figure}

Astrophysical observations of neutron stars provide another way to test EoSs. In Table~\ref{tab:properties} we report the radius of a compact star of mass $1.4\,\text{M}_\odot$ ($R_{1.4\Msolar}$), the maximum mass predicted by the EoSs ($M_{\text{max}}$), and the tidal deformability of a $1.4\,\text{M}_\odot$ star ($\Lambda_{1.4\Msolar}$). These values are consistent with the highest and most precise maximum mass measurement \cite{Riley:2021pdl}, combined NICER and XMM-Newton multimessenger constraints \cite{Pang:2021jta}, and tidal deformabilities consistent with current observations \cite{Miller:2021qha,Pang:2022rzc}. There may be non-trivial model dependence present in the multi-messenger constraints from Ref.~\cite{Pang:2021jta}. However, the mass-radius contours from NICER \cite{Riley:2019yda,Miller:2019cac,Riley:2021pdl,Miller:2021qha} are larger than the multi-messenger constraints. Given that our theories are consistent with the narrower multi-messenger constraints, they are also consistent with the NICER constraints. We plot the mass-radius curves for our four EoSs in Fig.~\ref{fig:mr_curve}. 
    
\begin{table*}
\begin{tabular}{lcccccccccc}
\hline
& $R_{1.4\Msolar}$ & $M_{\text{max}}$ & $\Lambda_{1.4\Msolar}$ & $M_*(\nsat)/m_N$ & $\mu_B^{\text{trans}}$ & $n_B^{\text{trans}}$ \\[-0.3ex]
    & [km] & [$\Msolar$] & & & [MeV] & [$\text{fm}^{-3}$]\\ 
\hline
QMC-RMF1 & 11.81 & 1.95 & 313 & 0.790 & 950 & 0.049\\
QMC-RMF2 & 11.94 & 2.04 & 357 & 0.769 & 950 & 0.045\\
QMC-RMF3 & 12.21 & 2.15 & 387 & 0.739 & 950 & 0.049\\
QMC-RMF4 & 12.27 & 2.21 & 470 & 0.730 & 950 & 0.043\\

Inference/observation
& $12.45^{+0.65}_{-0.65}$ & $2.072^{+0.066}_{-0.066}$ & $190^{+390}_{-120}$ & \\
\hline
\end{tabular}
\caption{Selected properties of our four EoSs in $\beta$-equilibrium at zero temperature. Inferred ranges for the neutron star radius come from Ref.~\cite{Miller:2021qha}, the maximum mass from Ref.~\cite{Riley:2021pdl}, and the tidal deformability from Ref.~\cite{PhysRevLett.121.161101}. The Dirac effective mass is consistent with values found by other microscopic calculations \cite{Bodmer:1989hdx,Margueron:2017eqc,Li:2018lpy,Chen:2014mza}. 
The transition density refers to the density in the homogeneous QMC-RMF phase.}
\label{tab:properties}
\end{table*}

\section{Conclusion}
We extend and improve our calculations of the four RMFTs developed in Ref.~\cite{Alford:2022bpp}. These RMFTs were obtained by simultaneously fitting the coupling constants of the RMFT to properties of pure neutron matter from $\chipt$, saturation properties of isospin-symmetric nuclear matter, and astrophysical observations of neutron star structure. We create three-dimensional tables of dense neutron, proton, and electron matter with temperature, baryon density, and charge fraction as independent variables over a range relevant for neutron star merger simulations. To create a table down to densities where our assumption of homogeneous nuclear matter breaks down, we combine our QMC-RMFs with the HS(IUF) EoS, which self-consistently models inhomogeneous nuclear matter at low densities and was fitted to properties of isospin-symmetric nuclear matter and finite nuclei, and therefore performs especially well in the parameter space near $y_q=0.5$.  We combine the QMC-RMF models in a thermodynamically consistent way with the HS(IUF) EoS. Each model is thermodynamically favored in the parameter range where it is designed to work best (QMC-RMFs near $y_q=0$ and HS(IUF) near $y_q=0.5$). Furthermore, we show that in parts of the parameter space where the temperature is high and the charge fraction is small, positrons and anti-nucleons (especially anti-protons) make a significant difference to the chemical potentials; this difference is especially important for transport properties. Therefore, we take the anti-particles of all particles in the model consistently into account, while they are often omitted in similar microscopic EoSs. The four EoS tables we construct are available on the CompOSE database \href{https://compose.obspm.fr/eos/297}{compose.obspm.fr/eos/297} \cite{qmc_rmf1,qmc_rmf2,qmc_rmf3,qmc_rmf4} and on GitLab at \href{https://gitlab.com/ahaber/qmc-rmf-tables}{gitlab.com/ahaber/qmc-rmf-tables}. These four EoS tables obey current observational, experimental, and theoretical constraints on dense matter and are ready to be used in simulations of neutron star mergers and other astrophysical applications.

\section{Acknowledgements}
We thank Elias R. Most for his input. This research was partly supported by the U.S. Department of Energy, Office of Science, Office of Nuclear Physics, under Award No.~\#DE-FG02-05ER41375.
The work of I.T. was supported by the U.S. Department of Energy, Office of Science, Office of Nuclear Physics, under contract No.~DE-AC52-06NA25396, and by the Laboratory Directed Research and Development program of Los Alamos National Laboratory under project number 20230315ER.

\clearpage
\bibliography{finite_t_chiPT_RMF}

\end{document}